# Consistency types for replicated data in a higher-order distributed programming language

Xin Zhao[a] and Philipp Haller[a]

a  KTH Royal Institute of Technology, Stockholm, Sweden

**Abstract**   Distributed systems address the increasing demand for fast access to resources and fault tolerance for data. However, due to scalability requirements, software developers need to trade consistency for performance. For certain data, consistency guarantees may be weakened if application correctness is unaffected. In contrast, data flow from data with weak consistency to data with strong consistency requirements is problematic, since application correctness may be broken.

In this paper, we propose lattice-based consistency types for replicated data (CTRD), a higher-order static consistency-typed language with replicated data types. The type system of CTRD supports shared data among multiple clients, and statically enforces noninterference between data types with weaker consistency and data types with stronger consistency. The language can be applied to many distributed applications and guarantees that updates of weakly-consistent data can never affect strongly-consistent data. We also extend the basic CTRD with an optimization that reduces synchronization for generating reference graphs.

## The Art, Science, and Engineering of Programming

**Consistency types for replicated data**

# 1 Introduction

Distributed systems are popular and used in many application domains. According to the CAP theorem [7], a fundamental result in distributed algorithms, all distributed applications must trade consistency for availability if network partitions must be tolerated. For this reason, different applications have different consistency requirements. For example, a banking service needs to ensure that users receive globally consistent information wherever they are, while a Twitter-like micro-blogging service needs to ensure the causal sequence of messages ("tweets") among related users. In order to provide higher performance, consistency requirements within the same application may also vary in different scenarios. For example, an online shopping service should provide product information with high availability (to optimize user experience), while the same service needs to ensure a safe and consistent transaction when a user completes their payment at checkout.

Much consideration is required when a developer fixes bugs or optimizes a distributed application. Unlike a local application, the verification work for a distributed program involves more expert experience and knowledge. Holt, Bornholt, Zhang, Ports, Oskin, and Ceze [10] show a common programming mistake caused by the misuse of consistency. They argue that an approach based on type checking could avoid similar errors. The authors introduce the so-called Inconsistent, Performance-bound, Approximate (IPA) storage system which makes consistency properties explicit for distributed data and which uses the type system to enforce consistency safety. However, their approach cannot prevent unsafe implicit data flow and no formal account of the type system is provided.

In this paper, we focus on the theoretical aspects of designing a type system for consistency safety. Importantly, we build on ideas from language-based information-flow security (which is novel in contrast to the work of Disciplined Inconsistency [10]) to enforce an essential noninterference property: in a well-typed program, weakly consistent data cannot flow into strongly consistent data. Moreover, we provide proofs of correctness properties such as type soundness and consistency guarantees.

This paper makes the following contributions:

- We introduce CTRD (Consistency Types for Replicated Data), a higher-order static consistency-typed language with identified references. The language supports shared data for multiple distributed clients, and distinguishes different system behaviors between weakly consistent and strongly consistent data.
- We prove type soundness for CTRD. We also prove that the type system guarantees noninterference, enabling the safe use of both weakly and strongly consistent data within the same program. Additionally, we prove consistency guarantees of the language, including sequential consistency for a certain class of operations.
- We introduce CTRD$^c$, a language extension for CTRD with records and clone operations. CTRD$^c$ enables safe propagation of local reference graphs from clients to replicated servers under strong consistency.

```
1  def numItemsAvailable(productId: Long): Int = {
2    getRef(productId).fastRead()
3  }
4  def realtimeDisplay(productId: Long) = {
5    display(numItemsAvailable(productId) + " in stock")
6  }
```

**(a)** Original display implementation

```
1  def checkOut(order: List[(productId: Long, num: Int)]) =
2  {
3    var errorFlag = false
4    var remaining = NUM_MAX
5    order foreach { i =>
6      var remaining = numItemsAvailable(i.productId)
7      if (remaining > i.num)
8        lockItems(i.productId, i.num, timer)
9      else errorFlag = true
10   }
11   if (!errorFlag) paymentProcess(order, timer)
12   else display("Please adjust number!")
13 }
```

**(b)** Extended implementation with error

**Figure 1** Shopping cart implementation

## 2 Motivating Example

In this section, we motivate our language design using a concrete example.

A typical program error that a distributed application developer might make is to mix the usage of data from different consistency levels. In the following example, we show a possible error occurring in a simplified shopping platform.

In figure 1a, the numItemsAvailable function applies a fastRead operation to get a weakly consistent remaining number of items of the required product from a random available server, so that the realtimeDisplay function provides fast access for users to get a better browsing experience.

During the development of the platform, the numItemsAvailable function might be reused by accident for implementing a checkOut function, as shown in figure 1b, which compares the current stock with the order number from a customer, and processes the order according to the availability. The reuse of the numItemsAvailable function does not cause any compilation errors; however, note that on line 8 of figure 1b, a weakly-consistent value returned from function numItemsAvailable is assigned to a variable remaining, which subsequently decides whether the final payment can be finished or not (lines 10-12). This problem arises easily, because (1) there might be two different programmers responsible for implementing the display and check-out parts, so they might fail to notice the problem of data inconsistency in different functions, and (2) it is difficult to reproduce and debug this kind of bug.

Analyzing the example in figure 1b, we note that the main issue is that the weakly consistent variable remaining affects the consistent payment operation. In order to rule out this unwanted dependency, we can apply information-flow techniques. There are two kinds of information-flows: direct (or explicit) and indirect (or implicit) flows.

For direct flow by direct assignment, a consistent variable should not be assigned available data. This dataflow constraint is also addressed in Disciplined Inconsistency [10]. For indirect flow by indirect influence such as conditional statements, a

**Consistency types for replicated data**

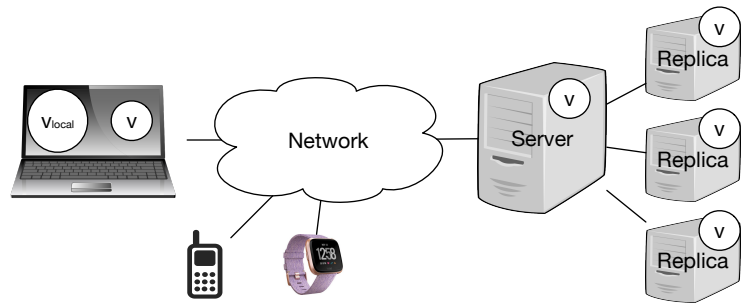

**Figure 2** Distributed system abstraction we focus on in the paper

condition with available data should not affect the consistent state, which cannot be detected by the type system of Disciplined Inconsistency [10].

Since many of the most widely-used applications are distributed systems with a mixed level of consistency, problems similar to the motivating example are hard to avoid. Inspired by techniques in information-flow security, which have been used to ensure confidentiality and integrity of information, we present a core language and type system, called CTRD, which prevents data flows that could break consistency properties.

## 3 Overview of CTRD

CTRD (Consistency Types for Replicated Data) is a higher-order static consistency-typed language with references. Its type system enables a *provably safe use* of weakly consistent data and strongly consistent data within the same application.

In this section, we first introduce the language abstraction we used in our system. We show how to fix the problem in the motivating example using CTRD's type system, and finally we discuss the features of CTRD and differences with closely related work.

### 3.1 Preliminaries

**Language abstraction for distributed systems**  A distributed system that we assume in the paper is given in figure 2. A location is a physical space where data is allocated, and the data is denoted "*v*". One can not only operate (i.e., create, read, or write) local data but also replicated data through different commands. The remote data are stored in multiple replicas for fault tolerance. The concept of locations will be used in the formalization of CTRD, and a location can be generated by allocating a new remote reference (see section 4).

When a client reads data from the server-side, if the data is always synchronized with the server and its corresponding replicas (i.e., the ones that are connected), then we say the data is strongly consistent. Otherwise, we say the data is weakly consistent, in other words, highly available since it only needs to communicate with a single replica.

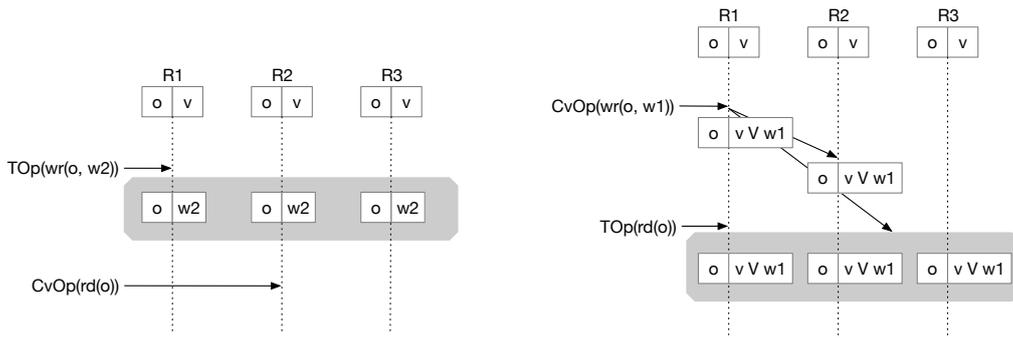

**(a)** Write in TOp manner and read in CvOp manner   **(b)** Write in CvOp manner and read in TOp manner

**Figure 3** Examples of different operations' behaviours in OACP

**Observable Atomic Consistency**   Zhao and Haller [29, 31] introduce the concept of observable atomic consistency (OAC), which allows fast updates on consistent objects as long as the updates are commutative, while still keeping strong consistency for totally ordered operations. Here we give a simple example to show some basic concepts and the corresponding protocol that preserves OAC. Figure 3 shows the different behaviors of different operations in the observable atomic consistency protocol (OACP).

We assume that there are three replicas in the distributed system for simplicity, all the values are part of a lattice so that they can be merged. In OACP, two main categories of operations are used. One is called "CvOp", which is commutative, and the system processes it asynchronously. The other one is called "TOp" which is totally ordered, and the system processes it as a single machine, which means all the replicas update at once through strong synchronization (i.e., using a protocol for distributed consensus like Paxos [14]).

In figure 3a, the initial value in location $o$ is $v$. When writing in a TOp manner, the location $o$ on all the replicas are updated to $w_2$ at the same time. When reading in a CvOp manner, one can get a return value as long as one of the replicas is reachable.

In figure 3b, when writing in a CvOp manner, the system updates the location $o$ asynchronously, and the value $w_2$ will be merged into the initial value $v$. When reading in a TOp manner, even though the asynchronous update might not be propagated to all the replicas, all the replicas will reach the most up-to-date value and then return this value to the client.

It is worth mentioning that if all the operations are CvOps, then the system provides eventual consistency. Moreover, if all the operations are TOps, then the system is strongly consistent.

## 3.2 Example solution in CTRD

Coming back to the motivating example in section 2, we notice that it is important to distinguish between highly available data and strongly consistent data. Based on this observation, in figure 4a, we introduce two types of operations `StrongRead` and `FastRead` for accessing the same remote location. The text in figure 4b explains the

**Consistency types for replicated data**

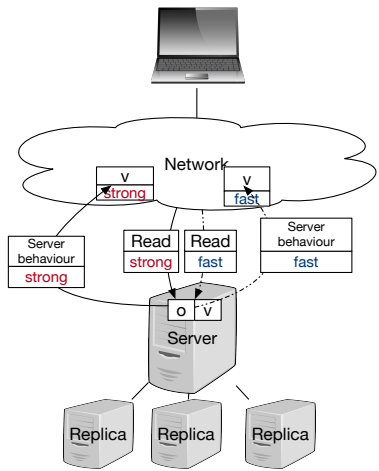

(a) Illustration of our solution

Step 1: Attach labels such as "strong" or "fast" to distinguish operations, return values and server behaviours.

Step 2: Make sure that the naming and the system behavior matches (i.e., StrongRead returns data that are strongly synchronized on the server side, and FastRead returns data as fast as possible.).

Step 3: Make sure that the client processes the return values based on rules that prevents "fast" values from affecting "strong" values neither in direct nor indirect ways.

(b) Step-by-step explanation of the solution

**Figure 4** One possible solution for example in section 2

general steps to solve the problem. Now, we introduce how to use CTRD to implement the idea.

**Step 1: Define attached labels** The type system of CTRD provides three kinds of replicated data types: (a) *consistent types* provide strong consistency; (b) *available types* provide availability, i.e., operations on available types never block; (c) *observable atomic consistent types* provide on-demand strong consistency.

Thus, including the local data types, we introduce four labels in CTRD: @loc, @con, @oac and @ava to represent local, consistent, observable atomic consistent and available types, respectively. Besides types, the labels are also related to values, operations, and system properties. @loc are for local values and operations. @con, @ava and @oac are for specified system properties. @con and @ava are also used for return values from the server-side and distributed operations (e.g., StrongRead and FastRead in figure 4a).

**Step 2: Identify proper consistency models** Consistent types are replicated data types which guarantee the consistency of replicas at all times. For this, a protocol for *distributed consensus*, such as Paxos [14] or Raft [22], is used to establish consistency upon each update of the data. As a result, operations applied to a consistent type behave as if the data was located and accessed on a single machine.

To achieve availability, operations on available types must weaken consistency. Specifically, in our language, available types provide *strong eventual consistency*, well-known from conflict-free replicated data types (CRDTs) [24]. CRDTs provide availability through asynchronous propagation of updates; the (eventual) convergence of replicas is ensured by commutativity properties of operations on their underlying data values.

▪ **Listing 1** Annotated example of figure 1b using CTRD labels

```
1  def numItemsAvailable(productId: Long): @ava Int = {
2      getRef(productId).FlexRead@ava()
3  }
4  def checkOut(order: List[productId: Long, num: Int]) = {
5      var errorFlag = false
6      var remaining = NUM_MAX
7      order foreach { i =>
8          var remaining = numItemsAvailable(i.productId) //@ava-labeled
9          if (remaining > i.num) //@ava-labeled condition
10             lockItems(i.productId, i.num, timer)
11         else errorFlag = true
12     }
13     if (!errorFlag) paymentProcess(order, timer)
14     else display("Please adjust number!")
15 }
```

Observable atomic consistent types provide flexible choices based on the application requirements. There are consistency models that provide a mixed level of consistency, such as fork consistency [17, 21], lazy replication [12], and red-blue consistency [16]. In our work, we use an *observable atomic consistency* protocol [29, 31] to establish consistency upon each totally-ordered update of the data.

The consistency level of OAC is stronger than eventual consistency and weaker than strong consistency. Thus, we can infer the lattice relationship that is introduced in the following section. The label oac we introduced before can represent a remote location with that property.

**Step 3: Apply information-flow techniques** Now we use the labels introduced above to annotate the example in figure 1b and get listing 1. The operation within function numItemsAvailable as well as the return value are labeled as @ava.

On line 8 of listing 1, the intention of the variable remaining is to get the current accurate remaining number of a product, but it is assigned as the return value of the numItemsAvailable function. If we can detect it is illegal to assign an available-typed value to a consistent-typed location, we would prevent the direct flow of data from available to consistent variables. From the information-flow point of view, label ava should be higher than label con. A complete partial order among labels is defined later in section 4 definition 4.1.

To achieve the goal of having no information-flow from available to consistent data, we also need to prevent *implicit flows*, which are caused by control flow. In listing 1, if we accept that remaining is by accident assigned to an ava-labeled variable, then the condition remaining > i.num has type available Boolean; thus, the consistent payment process depends on available variable remaining. CTRD avoids this situation by checking the label of the condition. If the condition has label @ava, the type system disallows the then-branch to execute terms that might mutate the state of a reference with a lower label, which in CTRD corresponds to a higher consistency level.

**Consistency types for replicated data**

We give one possible solution for the shopping cart example in the following. All we need to do is to implement a new function newNumItemsAvailable that applies FlexRead in a consistent manner and returns a con-labeled value, and to use this new function in the CHECKOUT function. In this case, the reference of productId is an oac-labeled location that can provide different labeled return values according to different operations.

```
1  def newNumItemsAvailable(productId: Long): @con Int = {
2      getRef(productId).FlexRead@con() }
```

### 3.3 Comparison with existing systems

**Comparison with the IPA storage system**   The work of Holt, Bornholt, Zhang, Ports, Oskin, and Ceze [10] has quantified analysis on error bounds and latency, but their work only enforces direct information-flow. In our work, we also prevent implicit information-flow, as shown in the example in figure 1(d). Moreover, they provide a prototype of a type system without operational semantics and soundness proof, which does not provide evidence that their type system is sound.

**Comparison with previous work in information-flow security**   Type systems in information-flow security such as $SSL_\mathcal{R}$ [26] carry a security label with a type to prevent data flow from high-security data to public observers. In CTRD, labels are used not only for tracking the usage of data from different consistency levels, but labels also fundamentally affect the operational semantics, and even data consistency. The language enforces safe data flow in an asynchronous setting (via async messages). The language provides not only noninterference but also properties such as a sequential consistency guarantee for con-labeled operations (theorem 5.1) and an eventual consistency guarantee for ava-labeled operations (theorem 5.2).

**Language features of CTRD**   The following paragraph summarizes the main features of CTRD.
1. CTRD is a distributed language and enforces safe data flow in an asynchronous setting.
2. CTRD binds runtime-generated locations to identifiers enabling multiple clients to access shared data in a distributed manner.
3. CTRD uses labels to annotate different consistency levels for information-flow tracking, which also fundamentally affects the operational semantics, and even data consistency.
4. CTRD uses labels to control the behavior of the system. (See the dynamic semantics in section 4.3 for more details.)

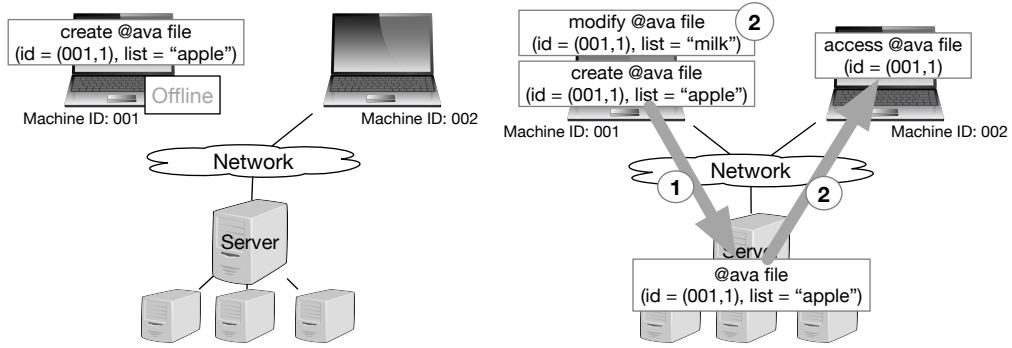

**(a)** Scenario: create an offline available file   **(b)** Scenario: access a shared file among clients

**Figure 5** An example to illustrate the reason for including await term

## 4 Formal Semantics

In this section, we formally introduce our consistency types for replicated data (CTRD). We use values in a lattice as one of the base values, and they are commonly used for achieving eventual consistency in distributed systems (e.g., CRDTs [24]).

### 4.1 Syntax

Figure 6 shows the syntax of our core language. The term language is essentially a typed functional language with references. The term $t[\ell]$ restricts the consistency level of term $t$. We introduce the identifier $id$ to bind and annotate locations such as $\text{ref}_\ell\ (t, id)$. In classical lambda calculus, a reference term creates a location in the local storage. Due to the distributed setup of CTRD, a reference with a distributed label $\ell$ (where $\ell$ can be con, oac or ava) creates a remote location storing term $t$ on the server-side with a given identifier $id$.

The term $\text{await}(id)$ is used to look up the bound location of an identifier. We use the example in figure 5 to explain why we need to introduce the await term. In figure 5a, a machine with ID 001 creates an available offline file which contains a mergeable list. Once the network is recovered, in figure 5b, the new file will appear on the server side as "1" shows, and the other machine with ID 002 can access the file with the file id and at the same time the machine 001 can update the context of the file (as "2" shows). This example shows a common pattern of many applications such as shared to-do list and collaborative editing. However, the behavior of machine 002 dynamically depends on the visibility of file id (001,1). Static analysis cannot capture the dynamic generation of $id$ but if we keep a global view of the generated $id$s and allow the program to wait for the $id$s to propagate to the server side, we can still use the technique to analyze the property of an execution history. Thus, the await term is necessary in our language.

We define dereference ($!t$) and assignment ($t := t$) as usual and additionally include two new terms $\text{FlexRead}_\ell(t)$ and $\text{FlexWrite}_\ell(t)$ for processing reads and writes with

**Consistency types for replicated data**

$$
\begin{array}{rcll}
\ell & ::= & \text{loc} \mid \text{con} \mid \text{oac} \mid \text{ava} & \text{label} \\
t & ::= & x \mid v \mid t[\ell] \mid t \bigoplus t \mid t \text{ op } t \mid t\, t \mid \text{if } x \text{ then } \{\, s\, \} \text{ else } \{\, t\, \} \\
& & \mid \text{ref}_\ell(t, id) \mid \text{await}(id) \mid !t \mid t := t \\
& & \mid \text{FlexRead}_\ell(t) \mid \text{FlexWrite}_\ell(t, t) & \text{terms} \\
r & ::= & d \mid \text{true} \mid \text{false} \mid (\lambda^\ell x : \tau.\, t) \mid \text{unit} \mid \boxed{o} & \text{raw value} \\
id & ::= & (\ell, n) \text{ where } n \in \mathbb{N} & \text{identifier} \\
v & ::= & r_\ell \mid \boxed{\text{duplicated}(t)} & \text{labeled value} \\
\tau & ::= & \text{Bool}_\ell \mid \text{Unit}_\ell \mid \text{Lat}_\ell \mid \text{Ref}_\ell\, \tau \mid \tau \xrightarrow{\ell}_\ell \tau & \text{types} \\
\bigoplus & ::= & \vee \mid \wedge & \text{meet and join} \\
\text{op} & ::= & \leq \mid < & \text{order operations}
\end{array}
$$

■ **Figure 6** Syntax of CTRD core language

oac-labeled locations. FlexRead$_{\text{ava}}$($t$) refers to "read in an available manner" where the return value does not need to be up-to-date. FlexRead$_{\text{con}}$($t$) refers to "read in a consistent manner" where strong synchronization among the servers is required before returning a value. FlexWrite$_{\text{ava}}$($t$) is to write in an available way where the updates are propagated asynchronously. While FlexWrite$_{\text{con}}$($t$) is to write to all the servers at the same time.

The term $d$ represents a value that is part of a lattice. The highlighted term $o$ and term duplicated($t$) only appear during the evaluation of a ref expression. $o$ is a reference location that is dynamically generated, and the term duplicated($t$) is to capture the situation where an identifier is already taken before a client creates a reference. The introduced identifier $id$ allows other clients to share the access to the same remote reference.

Another additional feature of the language is that each value, type constructor and even operation is annotated with a label $\ell$ to distinguish different consistency levels. Labels $\ell$ form a partial order $\leq$, where loc, con, oac and ava refer to labels that are attached to local, consistent, observable atomic consistent and available data types, respectively. The relations among them are defined in definition 4.1.

Function abstraction and arrow types are carrying a latent label [9] ($\ell$ besides $\lambda$ and $\ell$ above the arrow), which restricts the consistency level of the values that might be written during the execution of the function. In other words, a function with a lower latent label cannot be called in a higher labeled context. Recall the example in Section 2 figure 1, display and paymentProcess can be defined as an ava-labeled and a con-labeled function, respectively.

**Definition 4.1** (Partial order on consistency type labels). *The partial order of consistency type labels is:* loc $\leq$ con    con $\leq$ oac    oac $\leq$ ava.

Definition 4.1 is given according to our understanding of different consistency levels. Lower labels relate to higher consistency levels so that we can make use of the information-flow property.

$$\frac{T \in \{\text{Bool}, \text{Unit}, \text{Lat}\} \quad \ell \le \ell'}{T_\ell \le T_{\ell'}}$$

$$\frac{\tau_1' \le \tau_1 \quad \tau_2 \le \tau_2' \quad \ell_1 \le \ell_1' \quad \ell_2' \le \ell_2}{\tau_1 \xrightarrow{\ell_2}_{\ell_1} \tau_2 \le \tau_1' \xrightarrow{\ell_2'}_{\ell_1'} \tau_2'}$$

■ **Figure 7** Subtyping relation ≤ on types

$$\frac{\begin{array}{c}\Gamma; \Sigma; \Lambda_T; \ell_c \vdash t_1 : \text{Ref}_\ell \tau \quad \Gamma; \Sigma; \Lambda_T; \ell_c \vdash t_2 : \tau' \quad \tau' \le \tau \quad \ell_c \le \ell \\ \ell' = \text{label}(\tau') \quad \ell \ne \text{oac} \quad \ell' \ne \text{oac}\end{array}}{\Gamma; \Sigma; \Lambda_T; \ell_c \vdash t_1 := t_2 : \text{Unit}_\ell} \text{(T-Assign)}$$

■ **Figure 8** The typing rule for assignment statements

## 4.2 Static semantics

The typing judgement in CTRD is as follows: $\Gamma; \Sigma; \Lambda_T; \ell_c \vdash t : \tau$, which says that term $t$ has type $\tau$ under type environment $\Gamma$, store typing $\Sigma$, identifier typing $\Lambda_T$ and effect $\ell_c$. The type environment $\Gamma$ is a finite mapping from variables to their types. The store typing $\Sigma$ maps reference locations $o$ to the base type of the corresponding reference. For example, if $\Sigma(o) = \text{Bool}_\ell$ then $o$ has type $\text{Ref}_\ell \text{Bool}_\ell$. Similarly, the identifier typing $\Lambda_T$ maps reference identifiers to the type of the corresponding reference. The consistency label $\ell_c$ denotes the current consistency label (can also be called "effect" in a type-and-effect system) of the evaluation context for the given term, which prevents low-consistency terms from mutating the state of high-consistency references.

Subtyping rules are related to labels that are attached to the types. The definition is given in figure 7.

Now let us first illustrate the typing rule T-Assign in figure 8 to clarify the usage of labels and subtyping relations. In T-Assign, a term of type $\tau'$ is assigned to a reference of type $\tau$ only if $\tau'$ is a subtype of $\tau$. We can infer that $\tau'$ has a lower label than $\tau$ according to the subtyping rules. Based on this rule, it would be type-correct to assign a con value to an ava reference as long as their base types are equal. The oac label is excluded since we have an additional rule T-FlexWrt for specifying the situation.

The current consistency effect $\ell_c$ ensures that there are no illegal implicit flows. For example, $\Gamma; \Sigma; \Lambda_T; \text{ava} \vdash t_{\text{con}} := t'_{\text{con}}$ is illegal according to T-Assign, because $\ell_c = \text{ava} \not\le \ell = \text{con}$. Thus, when the current consistency effect is ava, we cannot mutate a con-labeled reference.

Figure 9 shows a few selected typing rules. In general, a CTRD source program $t$ is well-typed if $\cdot; \cdot; \cdot; \cdot \vdash t : \tau$.

The first few rules are similar to standard typing rules used in information-flow analysis.

T-RelOp types relational operations between two lattice values, yielding a boolean result after partial order comparison. ∨ is a join operation to compute the least upper bound of two labels.

**Consistency types for replicated data**

$$\frac{\Gamma; \Sigma; \Lambda_T; \ell_c \vdash t_i : \mathsf{Lat}_{\ell_i}}{\Gamma; \Sigma; \Lambda_T; \ell_c \vdash t_1 \text{ op } t_2 : \mathsf{Bool}_{\ell_1 \vee \ell_2}} \text{(T-RelOp)} \qquad \frac{\Gamma, x : \tau_1; \Sigma; \ell' \vdash t : \tau_2}{\Gamma; \Sigma; \Lambda_T; \ell_c \vdash (\lambda^{\ell'} x : \tau_1. t)_\ell : \tau_1 \xrightarrow{\ell'}_\ell \tau_2} \text{(T-Abs)}$$

$$\frac{\Gamma; \Sigma; \Lambda_T; \ell_c \vdash t_1 : \tau_{11} \xrightarrow{\ell'}_\ell \tau_{12} \quad \tau_2 \leq \tau_{11} \quad \Gamma; \Sigma; \Lambda_T; \ell_c \vdash t_2 : \tau_2 \quad \ell_c \vee \ell \leq \ell'}{\Gamma; \Sigma; \Lambda_T; \ell_c \vdash t_1 \, t_2 : \tau_{12} \vee \ell} \text{(T-App)}$$

$$\frac{\Gamma; \Sigma; \Lambda_T; \ell_c \vdash t : \mathsf{Bool}_\ell \quad \Gamma; \Sigma; \Lambda_T; \ell_c \vee \ell \vdash t_i : \tau_i \quad \tau = \tau_1 \vee \tau_2 \vee \ell}{\Gamma; \Sigma; \Lambda_T; \ell_c \vdash \textbf{if } t \textbf{ then } t_1 \textbf{ else } t_2 : \tau} \text{(T-If)}$$

$$\frac{\begin{array}{c}\ell \neq \mathsf{oac} \quad \Gamma; \Sigma; \Lambda_T; \ell_c \vdash t : \tau' \quad \mathsf{label}(\tau') \leq \ell \quad \ell_c \leq \ell \quad \tau = \tau' \vee \ell \\ \ell = \mathsf{ava} \implies \mathsf{raw}(\tau') = \mathsf{Lat} \quad \mathsf{label}(\tau') < \ell \implies \mathsf{refs}(t) = \emptyset \quad id.\ell = \ell\end{array}}{\Gamma; \Sigma; \Lambda_T; \ell_c \vdash \mathsf{ref}_\ell(t, id) : \mathsf{Ref}_\ell \, \tau} \text{(T-Ref-1)}$$

$$\frac{\begin{array}{c}\ell = \mathsf{oac} \quad \Gamma; \Sigma; \Lambda_T; \ell_c \vdash t : \tau' \quad \mathsf{label}(\tau') < \ell \quad \ell_c \leq \ell \quad \tau = \tau' \vee \ell \\ \mathsf{raw}(\tau') = \mathsf{Lat} \quad id.\ell = \ell\end{array}}{\Gamma; \Sigma; \Lambda_T; \ell_c \vdash \mathsf{ref}_\ell(t, id) : \mathsf{Ref}_\ell \, \tau} \text{(T-Ref-2)}$$

$$\frac{\Gamma; \Sigma; \Lambda_T; \ell_c \vdash t : \mathsf{Ref}_{\mathsf{oac}} \, \tau \quad \ell = \mathsf{con} \vee \mathsf{ava}}{\Gamma; \Sigma; \Lambda_T; \ell_c \vdash \mathsf{FlexRead}_\ell(t) : \mathsf{raw}(\tau) \vee \ell} \text{(T-FlexRd)} \qquad \frac{\Gamma; \Sigma; \Lambda_T; \ell_c \vdash t_1 : \mathsf{Ref}_{\mathsf{oac}} \, \tau \quad \Gamma; \Sigma; \Lambda_T; \ell_c \vdash t_2 : \tau \quad \ell = \mathsf{con} \vee \mathsf{ava}}{\Gamma; \Sigma; \Lambda_T; \ell_c \vdash \mathsf{FlexWrite}_\ell(t_1, t_2) : \mathsf{Unit}_\ell} \text{(T-FlexWrt)}$$

■ **Figure 9** Selected typing rules

T-Abs type checks the function body with the latent label $\ell'$, and the arrow type has the same label as the function abstraction.

T-App enforces consistency restrictions in a standard way. The latent label $\ell'$ is an upper bound for the current consistency level and the operator label. The $\vee$ between a type $\tau_{12}$ and a label $\ell$ is defined as an operation to join $\ell$ with the label of $\tau_{12}$. The type of the entire term $\tau_{12}$ joins the label $\ell$ to preserve the consistency level.

T-If checks the current type of the predicate and propagates the label to the type checking of each branch statement. In this way, the rule prevents implicit information-flow from higher labels to lower labels. For types that have the same raw type, $\vee$ is defined to compute a result type with the joined label (i.e., $\tau_\ell \vee \ell' = \tau_{\ell \vee \ell'}$).

Starting from T-Ref-1, there are additional attributes for supporting the distributed feature of CTRD.

T-Ref-1 applies when the label $\ell$ is not oac. It type checks the reference body $t$ and then compares its label, the consistency label $\ell$ of term ref, and the current consistency effect $\ell_c$ in a standard way. It also checks whether the label of $id$ is the same as the label of the reference. Besides that, this rule has two more restrictions. First, if $\ell$ is ava, then the base type of $t$ is a lattice-type. Second, if the label of $t$ is lower than $\ell$, then $t$ does not contain references (i.e., $\mathsf{refs}(v) = \emptyset$). The reasons for introducing these restrictions are: 1) for a reference with the label ava, it will create a remote location in an available manner (see figure 5a as an example). The requirement of having the stored value to be lattice-based is to simplify the solution for conflicts between concurrent updates. 2) when a ref term is type-checked, the consistency label $\ell$ is attached to the type of the reference body $t$. In a local setup, we do not need to consider the format of $t$. However, our labels are representing consistency behaviors, upgrading the label of the reference body means that the storage behavior needs to change for the stored value. For example, term $\mathsf{ref}_{\mathsf{con}}(\mathsf{ref}_{\mathsf{loc}}(3, id1), id2)$ needs to store a local reference distributedly, which is meaningless for remote servers since the local location is not accessible for them.

T-Ref-2 applies when the label $\ell$ is oac. Because the consistency level of on-demand strong consistent data is either consistent or available at a fixed point of time, data with oac is hard to map to the real world. Thus, we treat the oac label in a special way. Different from T-Ref-1, the label of $\tau'$ is lower than $\ell$ to exclude the condition where value $t$ has a oac label. (Note that $t$ cannot be a reference here due to the restriction that the raw type of $\tau'$ need to be lattice-based, and the reason is the same as in T-Ref-1.)

T-FlexRd applies to term $t$ with the oac label. The consistency label $\ell$ can only be con or ava, and the label of the result type is decided by $\ell$. T-FlexWrt applies to term $t_1$ with oac label. The second term $t_2$ is the value that is assigned to the first term, and the label of the term can only be con or loc. Similar to T-FlexRd, the label of the result type is decided by $\ell$, which can only be con or ava.

Other typing rules such as T-Deref are defined in a standard way, thus we omit them; they appear in the companion technical report [30].

### 4.3 Dynamic semantics

We formalize the dynamic semantics as a small-step operational semantics based on two judgements:

$$t_1 \mid \mu_1 \mid b_1 \mid \lambda_1 \xrightarrow{a}{}^i t_2 \mid \mu_2 \mid b_2 \mid \lambda_2$$

and

$$\{\langle t_1 \mid \mu_1 \mid b_1 \mid \lambda_1 \rangle^i\} \cup P \mid M_1 \mid S_1 \mid \Lambda_1 \xrightarrow{a}{\twoheadrightarrow} \{\langle t_2 \mid \mu_2 \mid b_2 \mid \lambda_2 \rangle^i\} \cup P \mid M_2 \mid S_2 \mid \Lambda_2$$

The first judgement says that a term $t_1$, a local store $\mu_1$, a message buffer $b_1$ and a local mapping from identifiers to locations $\lambda_1$ reduce to $t_2$, $\mu_2$, $b_2$ and $\lambda_2$, respectively, with an action $a$. The second judgement reduces the entire cloud state [2] which contains a set of client programs $P$, a multiset of messages $M$, a set of servers $S$, and a global mapping from identifiers to locations $\Lambda$. An action $a$ is a pair $(\ell_c, \mathsf{op})$ which contains a current consistency label $\ell_c$ in the context and an operation op which is either rd, wr, ref, or $\varepsilon$. op is used for analysing the consistency guarantees for CTRD in section 5.2, so we omit the discussion about it in this section.

**Local reduction**   Figure 10 shows a few selected local reduction rules $t_1 \mid \mu_1 \mid b_1 \mid \lambda_1 \xrightarrow{a}{}^i t_2 \mid \mu_2 \mid b_2 \mid \lambda_2$. The other rules are omitted and appear only in the companion technical report [30]. Local operations follow the general reduction rules. $b$ is a message

**Consistency types for replicated data**

$$\frac{t_1 \mid \mu_1 \mid b_1 \mid \lambda_1 \xrightarrow{a}^i t_2 \mid \mu_2 \mid b_2 \mid \lambda_2}{E[t_1] \mid \mu_1 \mid b_1 \mid \lambda_1 \xrightarrow{a}^i E[t_2] \mid \mu_2 \mid b_2 \mid \lambda_2} \quad \text{(E-Eval)}$$

$$\frac{d = d_1 \ [\oplus] \ d_2 \quad \ell = \ell_1 \vee \ell_2 \quad a = (\ell_c, \varepsilon)}{d_{1\ell_1} \oplus d_{2\ell_2} \mid \mu \mid b \mid \lambda \xrightarrow{a}^i d_\ell \mid \mu \mid b \mid \lambda} \quad \text{(E-LatOp)}$$

$$\frac{o \notin dom(\mu) \quad id \notin dom(\lambda) \quad \mu' = \mu[o \mapsto v \vee \ell_c] \quad \lambda' = \lambda[id \mapsto o] \quad a = (\ell_c, \varepsilon)}{\mathsf{ref}_{\mathsf{loc}}(v, id) \mid \mu \mid b \mid \lambda \xrightarrow{a}^i o_{\mathsf{loc}} \mid \mu' \mid b \mid \lambda'} \quad \text{(E-LocalRef)}$$

$$\frac{\begin{array}{c} o = (i, \iota) \text{ where } i \in Ids \text{ and } \iota \text{ fresh} \quad \mu' = \mu[o \mapsto v \vee \ell_c \vee \mathsf{ava}] \quad a = (\ell_c, \mathit{ref}^{\mathbb{v}}_{\mathsf{ava}}(o, v)) \\ id \notin dom(\lambda) \quad \lambda' = \lambda[id \mapsto o] \quad b' = b \cdot \mathsf{update}[o, id, v, i, \emptyset, \mathbb{v}] \quad \mathbb{v} \text{ fresh} \end{array}}{\mathsf{ref}_{\mathsf{ava}}(v, id) \mid \mu \mid b \mid \lambda \xrightarrow{a}^i o_{\mathsf{ava}} \mid \mu' \mid b' \mid \lambda'} \quad \text{(E-AvaRef)}$$

$$\frac{\ell = \mathsf{loc} \vee \mathsf{ava} \quad id \in dom(\lambda) \quad a = (\ell_c, \varepsilon)}{\mathsf{ref}_\ell(v, id) \mid \mu \mid b \mid \lambda \xrightarrow{a}^i \mathsf{duplicated}(\mathsf{ref}_\ell(v, id)) \mid \mu \mid b \mid \lambda} \quad \text{(E-Ref-Dup)}$$

$$\frac{o \in dom(\mu) \quad v = \mu(o) \quad b' = b \cdot \mathsf{req}[\lambda.getkey(o), i] \quad \mathbb{v} \text{ fresh} \quad a = (\ell_c, \mathit{rd}^{\mathbb{v}}_{\mathsf{ava}}(o, v, \mu))}{!o_{\mathsf{ava}} \mid \mu \mid b \mid \lambda \xrightarrow{a}^i v \vee \mathsf{ava} \mid \mu \mid b' \mid \lambda} \quad \text{(E-AvaDerefi)}$$

$$\frac{\begin{cases} \mathbf{if} \ o \in dom(\mu) \ \mathbf{then} \ v' = \mu(o) \vee v \quad b' = b \cdot \mathsf{update}[o, \lambda.getkey(o), v, i, \emptyset, \mathbb{v}] \\ \mathbf{else} \ v' = v \quad b' = b \cdot \mathsf{update}[o, \text{""}, v, i, \emptyset, \mathbb{v}] \end{cases} \\ \mu' = \mu[o \mapsto v' \vee \ell_c \vee \mathsf{ava}] \quad \mathbb{v} \text{ fresh} \quad a = (\ell_c, \mathit{wr}^{\mathbb{v}}_{\mathsf{ava}}(o, v))}{o_{\mathsf{ava}} := v \mid \mu \mid b \mid \lambda \xrightarrow{a}^i \mathsf{unit}_{\mathsf{ava}} \mid \mu' \mid b' \mid \lambda} \quad \text{(E-avaAssign)}$$

$$\frac{id \in dom(\lambda) \quad o = \lambda(id) \quad a = (\ell_c, \varepsilon)}{\mathsf{await}(id) \mid \mu \mid b \mid \lambda \xrightarrow{a}^i o \mid \mu \mid b \mid \lambda} \quad \text{(E-Awaiti)}$$

■ **Figure 10** CTRD: Local reduction

sending buffer for modeling the potential network delay or interruption in eventually consistent systems. $\lambda$ is a local mapping from identifiers to locations.

E-Eval defines the form of local reduction relations. We use evaluation contexts here. Each evaluation context is a term with a hole ([]) somewhere inside it. We write $E[t]$ for the term obtained by replacing the hole in $E$ with $t$. Evaluation contexts are defined as follows.

$$\begin{aligned} E \ ::= \ & [] \mid E \oplus t \mid v \oplus E \mid E \text{ op } t \mid v \text{ op } E \mid E \ t \mid v \ E \mid \mathbf{if} \ E \ \mathbf{then} \ t \ \mathbf{else} \ t \mid !E \mid E := t \\ & \mid v := E \mid \mathsf{ref}_\ell(E, id) \mid \mathsf{FlexRead}_\ell(E) \mid \mathsf{FlexWrite}_\ell(E, t) \mid \mathsf{FlexWrite}_\ell(v, E) \end{aligned}$$

E-LatOp is a simple rule that makes a reduction on terms without modifying store $\mu$, buffers $b$, or mapping $\lambda$. Some rules modify the store and mapping, for example, in E-LocalRef, when a term $\text{ref}^\tau_{\text{loc}}(v, id)$ needs to be reduced, it first generates a location $o$ that is not in the domain of the store $\mu$ ($o \notin dom(\mu)$). Then $\mu$ extends its mapping relation with a labeled $v$ associated to $o$. Moreover, $\lambda$ extends its mapping with the reference identifier $id$ to the location $o$.

E-AvaRef creates an available reference. In contrast to E-LocalRef, the generated location has a decentralized identifier which are defined as $(i, \iota)$ where $i$ is the program ID and $\iota$ is freshly generated. The value stored in the location is stamped with label ava, and since it requires communication but not instant creation on the server-side, a message of the form $\text{update}[o, id, v, i, \emptyset, \mathbb{v}]$ which contains the update information $(o, id, v, i)$, a set of receivers $\emptyset$ that already received the message and the event id $\mathbb{v}$ is stored in the buffer for subsequent, asynchronous processing. For E-LocalRef and E-AvaRef, we exclude the situation when $id \in dom(\lambda)$, and this is handled in E-Ref-Dup where a duplicated($t$) term is generated.

Available data types are only weakly consistent; thus, it is sufficient for E-AvaDeref1 to obtain the value directly from the local store if the location is in the domain of the local storage. Meanwhile, a message of the form $\text{req}[id, i]$ is stored in the buffer for requesting a newer state from an available server. (For the condition where the location is not in the domain of the local store, see E-AvaDeref2 in figure 11.) E-AvaAssign first updates the local store depending on whether the location is locally buffered or not, and then puts a message into the buffer for message propagation. E-ConRef, E-ConDeref and E-ConAssign involve network connection thus we will discuss them later in the distributed reduction.

When a run-time location is created, and one has an identifier associated to it, E-Await1 can return a location if the provided $id$ exists in the domain of the local $\lambda$. (For the condition where the $id$ is out of the scope of the local $\lambda$, see E-Await2 in figure 11.)

All expressions propagate the current program label $\ell_c$ to subterms, and there is no additional runtime checking for types and consistency labels required.

**Distributed reduction**  Figure 11 and figure 12 show the selected distributed reduction rules (see the companion technical report [30] for the omitted rules).

$$\{\langle t_1 \mid \mu_1 \mid b_1 \mid \lambda_1 \rangle^i\} \cup P \mid M_1 \mid S_1 \mid \Lambda_1 \xrightarrow{a} \{\langle t_2 \mid \mu_2 \mid b_2 \mid \lambda_2 \rangle^i\} \cup P \mid M_2 \mid S_2 \mid \Lambda_2$$

Each client consists of a tuple $\langle t \mid \mu \mid b \mid \lambda \rangle$. $M$ is the abstraction for network connections (messages). Each server has a pair structure $(s, \text{seq})$ where $s$ is similar as local storage $\mu$ and seq is a sequence of events occurring on the server side which we use for analysing the consistency guarantees in section 5.2. Moreover, we maintain a separate global abstract store $\Lambda$ mapping identifiers to locations: $\Lambda(i) = o$ if identifier $i$ refers to a location $o$.

E-Local reveals the relationship between reduction relations $\rightarrow^i$ and $\twoheadrightarrow^i$. E-Await2 shows that the system will be blocked while waiting for the identifier $id$ to appear in the global mapping $\Lambda$, and continues to be reduced to a location $\Lambda(id)$. Local mapping $\lambda$ will also be updated accordingly. E-AvaDeref2 shows the behavior of the system

**Consistency types for replicated data**

$$\frac{id \notin dom(\lambda) \quad \Lambda(id) = o \quad \lambda' = \lambda[id \mapsto o] \quad a = (\ell_c, \varepsilon)}{\{\langle \text{await}(id) \mid \mu \mid b \mid \lambda \rangle^i\} \cup P \mid M \mid S \mid \Lambda \xrightarrow{a} \{\langle o \vee \ell \mid \mu \mid b \mid \lambda' \rangle^i\} \cup P \mid M \mid S \mid \Lambda} \quad \text{(E-Await2)}$$

$$\frac{o \notin dom(\mu) \quad \exists S_r \in S.\ S_r.s(o) = v \quad \mu' = \mu[o \mapsto v] \quad \mathbbm{v}\ \text{fresh} \quad a = (\ell_c, rd^{\mathbbm{v}}_{\text{ava}}(o, v, S_r))}{\{\langle !o_{\text{ava}} \mid \mu \mid b \mid \lambda \rangle^i\} \cup P \mid M \mid S \mid \Lambda \xrightarrow{a} \{\langle v \vee \text{ava} \mid \mu' \mid b \mid \lambda \rangle^i\} \cup P \mid M \mid S \mid \Lambda} \quad \text{(E-AvaDeref2)}$$

$$\frac{\begin{array}{c} id \notin dom(\Lambda) \quad o = (i, \iota)\ \text{where}\ \iota\ \text{fresh} \quad i \in \textit{Ids} \\ S' = \bigcup S'_r\ \text{where}\ \forall S_r \in S, S'_r = (S_r.s[o \mapsto v'], \mathbbm{v} \cdot S_r.seq) \\ \Lambda' = \Lambda[id \mapsto o] \quad v' = v \vee \ell_c \vee \ell \quad a = (\ell_c, \textit{ref}^{\mathbbm{v}}_{\ell}(o, v)) \quad \mathbbm{v}\ \text{fresh} \end{array}}{\{\langle \text{ref}_{\text{con}}(v, id) \mid \mu \mid b \mid \lambda \rangle^i\} \cup P \mid M \mid S \mid \Lambda \xrightarrow{a} \{\langle o_{\text{con}} \mid \mu \mid b \mid \lambda \rangle^i\} \cup P \mid M \mid S' \mid \Lambda'} \quad \text{(E-ConRef)}$$

$$\frac{\ell = \text{con} \vee \text{oac} \quad id \in dom(\Lambda) \quad a = (\ell_c, \varepsilon)}{\{\langle \text{ref}_{\ell}(v, id) \mid \mu \mid b \mid \lambda \rangle^i\} \cup P \mid M \mid S \mid \Lambda \xrightarrow{a} \{\langle \text{duplicated}(\text{ref}_{\ell}(v, id)) \mid \mu \mid b \mid \lambda \rangle^i\} \cup P \mid M \mid S \mid \Lambda} \quad \text{(E-ConRef-Dup)}$$

$$\frac{\exists S_r \in S.\ S_r.s(o) = v \quad \mathbbm{v}\ \text{fresh} \quad a = (\ell_c, rd^{\mathbbm{v}}_{\text{con}}(o, v, S_r))}{\{\langle !o_{\text{con}} \mid \mu \mid b \mid \lambda \rangle^i\} \cup P \mid M \mid S \mid \Lambda \xrightarrow{a} \{\langle v \vee \text{con} \mid \mu \mid b \mid \lambda \rangle^i\} \cup P \mid M \mid S \mid \Lambda} \quad \text{(E-ConDeref)}$$

$$\frac{\begin{array}{c} v' = v \vee \ell_c \vee \text{con} \quad \mathbbm{v}\ \text{fresh} \quad a = (\ell_c, wr^{\mathbbm{v}}_{\text{con}}(o, v)) \\ S' = \bigcup S'_r\ \text{where}\ \forall S_r \in S, S'_r = (S_r.s[o \mapsto v'], \mathbbm{v} \cdot S_r.seq) \end{array}}{\{\langle o_{\text{con}} := v \mid \mu \mid b \mid \lambda \rangle^i\} \cup P \mid M \mid S \mid \Lambda \xrightarrow{a} \{\langle \text{unit}_{\text{con}} \mid \mu \mid b \mid \lambda \rangle^i\} \cup P \mid M \mid S' \mid \Lambda} \quad \text{(E-ConAssign)}$$

**Figure 11** CTRD: Selected distributed reduction (1)

when the local store does not contain the required location $o$, which is a result returned from an available(accessible) server.

E-ConRef creates a reference in a consistent manner, and the generated location is stamped with a label con. The creation of the reference requires a totally-ordered update of all servers. We express it by simultaneously changing the state of all servers, in one step. A practical implementation would require an algorithm for distributed consensus. However, our semantics is designed for reasoning about source programs, on a high level, instead of the implementation of the underlying distributed algorithms. Therefore, we abstract from the underlying distributed consensus algorithm. E-ConRef-Dup is similar as E-Ref-Dup that captures the situation when $id \in dom(\Lambda)$.

E-ConDeref applies to locations with con labels. The rule returns a value from a random server, and the value is still consistent because all the write operations are strongly synchronized. Similarly, E-ConAssign applies to con-labeled locations assignment. The rule updates the remote side in one step to express the fact that to get a consistent result from the server-side or to assign a consistent data type, all the servers need to synchronize and reach a consistent state before completing the consistent operation.

$$\frac{\begin{array}{c} id \notin dom(\Lambda) \qquad o = (i, \iota) \text{ where } \iota \text{ fresh} \qquad i \in Ids \\ \lambda' = \lambda[id \mapsto o] \qquad v' = v \lor \ell_c \lor \ell \qquad \mu' = \mu[o \mapsto v'] \\ S' = \bigcup S'_r \text{ where } \forall S_r \in S, S'_r = (S_r.s[o \mapsto v'], \mathbbm{v} \cdot S_r.seq) \\ \Lambda' = \Lambda[id \mapsto o] \qquad a = (\ell_c, ref^{\mathbbm{v}}_\ell(o, v)) \qquad \mathbbm{v} \text{ fresh} \end{array}}{\{\langle \text{ref}_{\text{oac}}(v, id) \mid \mu \mid b \mid \lambda \rangle^i\} \cup P \mid M \mid S \mid \Lambda \xrightarrow{a} \{\langle o_{\text{oac}} \mid \mu' \mid b \mid \lambda' \rangle^i\} \cup P \mid M \mid S' \mid \Lambda'} \text{ (E-OacRef)}$$

$$\frac{\begin{array}{c} \begin{cases} \textbf{if } o \in dom(\mu) \textbf{ then } w = \mu(o), v' = (w \lor v) \lor \ell_c \lor \text{ava}, \mu' = \mu[o \mapsto v'] \\ \textbf{else } v' = v \lor \ell_c \lor \text{ava}, \mu' = \mu[o \mapsto v'] \end{cases} \\ b' = b \cdot \text{update}[o, \lambda.getkey(o), v, i, \emptyset] \qquad \mathbbm{v} \text{ fresh} \qquad a = (\ell_c, wr^{\mathbbm{v}}_{\text{con}}(o, v')) \end{array}}{\{\langle \text{FlexWrite}_{\text{ava}}(o, v) \mid \mu \mid b \mid \lambda \rangle^i\} \cup P \mid M \mid S \mid \Lambda \xrightarrow{a} \{\langle \text{unit}_{\text{ava}} \mid \mu' \mid b' \mid \lambda \rangle^i\} \cup P \mid M \mid S \mid \Lambda} \text{ (E-FlexWrt-Ava)}$$

$$\frac{\begin{array}{c} v' = v \lor \ell_c \lor \text{con} \qquad \mu' = \mu[o \mapsto v'] \qquad \mathbbm{v} \text{ fresh} \qquad a = (\ell_c, wr^{\mathbbm{v}}_{\text{con}}(o, v')) \\ S' = \bigcup S'_r \text{ where } \forall S_r \in S, S'_r = (S_r.s[o \mapsto v'], \mathbbm{v} \cdot S_r.seq) \end{array}}{\{\langle \text{FlexWrite}_{\text{con}}(o, v) \mid \mu \mid b \mid \lambda \rangle^i\} \cup P \mid M \mid S \mid \Lambda \xrightarrow{a} \{\langle \text{unit}_{\text{con}} \mid \mu' \mid b \mid \lambda \rangle^i\} \cup P \mid M \mid S' \mid \Lambda} \text{ (E-FlexWrt-Con)}$$

$$\frac{\begin{cases} \textbf{if } o \in dom(\mu) \textbf{ then } r_\ell = \mu(o), \mu' = \mu, a = (\ell_c, rd^{\mathbbm{v}}_{\text{ava}}(o, r_{\text{ava}}, \mu)) \\ \textbf{else } \exists S_r \in S.r_\ell = S_r(o), \mu' = \mu[o \mapsto r_\ell], a = (\ell_c, rd^{\mathbbm{v}}_{\text{ava}}(o, r_{\text{ava}}, S_r)) \end{cases}}{\{\langle \text{FlexRead}_{\text{ava}}(o) \mid \mu \mid b \mid \lambda \rangle^i\} \cup P \mid M \mid S \mid \Lambda \xrightarrow{a} \{\langle r_{\text{ava}} \mid \mu' \mid b \mid \lambda \rangle^i\} \cup P \mid M \mid S \mid \Lambda} \text{ (E-FlexRd-Ava)}$$

$$\frac{\begin{array}{c} v_{\text{con}} = \forall S_r \in S. \bigcup S_r.s(o) \qquad S' = \forall S_r \in S.S_r.s[o \mapsto v] \qquad \mu' = \mu[o \mapsto v_{\text{con}}] \\ a = (\ell_c, rd^{\mathbbm{v}}_{\text{con}}(o, v_{\text{con}}, S')) \end{array}}{\{\langle \text{FlexRead}_{\text{con}}(o) \mid \mu \mid b \mid \lambda \rangle^i\} \cup P \mid M \mid S \mid \Lambda \xrightarrow{a} \{\langle v_{\text{con}} \mid \mu' \mid b \mid \lambda \rangle^i\} \cup P \mid M \mid S' \mid \Lambda} \text{ (E-FlexRd-Con)}$$

■ **Figure 12** CTRD: Distributed reduction (2)

E-OacRef creates a reference with oac-label similarly as E-ConRef. The only difference is that the local storage is also updated due to the fact that available operations need to have local access support for getting fast responses.

E-FlexWrt-Ava and E-FlexWrt-Con perform similarly as E-AvaAssign and E-ConAssign, respectively, but specifically for modifying references with oac labels. When we write to a location $o$ in an available manner, message buffer $b$ is updated, and the message will be propagated in an asynchronized way. They allow the programmer to apply different consistency protocols to the same reference location.

E-FlexRd-Ava performs a fast read operation that returns either the local value or the value from an available server so that the consistency object can still be fast read if we want to improve performance. E-FlexRd-Con performs as a strong read operation which means one needs to force synchronization on the servers to reach a consistent state. Using the property of OACP protocol, we first merge the states from all the servers and then return a consistent value from the server side.

**Consistency types for replicated data**

$$\frac{b = m \cdot b' \qquad M' = m \cup M \qquad a = (\ell_c, \varepsilon)}{\{\langle t \mid \mu \mid b \mid \lambda \rangle^i\} \cup P \mid M \mid S \mid \Lambda \xrightarrow{a} \{\langle t \mid \mu \mid b' \mid \lambda \rangle^i\} \cup P \mid M' \mid S \mid \Lambda} \quad \text{(E-Send)}$$

$$\frac{M = \{\text{update}[o, id, v, i, R, \mathbb{v}]\} \uplus M' \qquad R = ids(S) \qquad a = (\ell_c, \varepsilon)}{P \mid M \mid S \mid \Lambda \xrightarrow{a} P \mid M' \mid S \mid \Lambda} \quad \text{(E-GC)}$$

$$\frac{\begin{array}{c} M = \{\text{req}[id, i]\} \cup M' \qquad o = \Lambda(id) \qquad \exists S_r \in S. S_r(o) = v \qquad o' = \lambda(id) \\ \mu' = \mu[o' \mapsto v \vee \ell_c] \qquad a = (\ell_c, \varepsilon) \end{array}}{\{\langle t \mid \mu \mid b \mid \lambda \rangle^i\} \cup P \mid M \mid S \mid \Lambda \xrightarrow{a} \{\langle t \mid \mu' \mid b \mid \lambda \rangle^i\} \cup P \mid M' \mid S \mid \Lambda} \quad \text{(E-Process-Request)}$$

$$\frac{\begin{array}{c} M = \{\text{update}[o, id, v, i, R, \mathbb{v}]\} \cup M'' \qquad S = \{S_r\} \uplus S'' \qquad r \notin R \\ \begin{cases} \textbf{if } id \notin dom(\Lambda) \textbf{ then } \Lambda' = \Lambda[id \mapsto o], o' = o \\ \textbf{else } o' = \Lambda(id) \end{cases} \\ \begin{cases} \textbf{if } o' \notin S_r \textbf{ then } S'_r = (S_r.s[o' \mapsto v \vee \ell_c], \mathbb{v} \cdot S_r.seq) \\ \textbf{else } S'_r = (S_r.s[o' \mapsto v \vee S_r.s(o') \vee \ell_c], \mathbb{v} \cdot S_r.seq) \end{cases} \quad S' = \{S'_r\} \cup S'' \\ M' = \{\text{update}[o, id, v, i, R \cup \{r\}, \mathbb{v}]\} \cup M'' \qquad a = (\ell_c, wr^{\mathbb{v}}_{\text{ava}}(o, v)) \end{array}}{P \mid M \mid S \mid \Lambda \xrightarrow{a} P \mid M' \mid S' \mid \Lambda'} \quad \text{(E-Process-Update)}$$

■ **Figure 13** CTRD: Message processing

Figure 13 shows the message processing of CTRD. E-Send moves a message from buffer $b$ to message set $M$. E-GC removes a message that has already been received by all servers. E-ProcessUpdate processes a message $\text{update}[o, id, v, i, R, \mathbb{v}]$ which contains the update information $(o, id, v, i)$, a set of receivers $R$ that already received the message and the event id $\mathbb{v}$. The rule updates a server that does not belong to $R$ and adds a new message for further propagation. If the message contains a location $o$ that does not exist on the server, then the server creates a new mapping; otherwise, the value at the same location is updated. E-ProcessRequest processes a message $\text{req}[id, i]$. The rule pushes the state of location $\Lambda(id)$ in one of the replicas back to client $i$, updating its local location $o'$.

**Well-formed configurations** Figure 14 shows a few selected well-formed configurations. The complete well-formed rules can be found in the companion technical report [30]. These rules are essential for establishing subject reduction (section 5.1). WF-Store defines the well typeness of the store with respect to a store typing $\Sigma$ and an identifier typing $\Lambda_T$. The remaining rules lift well-formedness to program configurations (WF-ProgramConfig), sets of client programs (WF-Program) and configurations (WF-Config), respectively.

$$\frac{dom(\mu) \subseteq dom(\Sigma) \quad \forall o \in dom(\mu) \, . \, \cdot; \Sigma; \cdot; \cdot \vdash \mu(o) : \tau \wedge \tau \leq \Sigma(o)}{\Sigma; \Lambda_T \vdash \mu} \quad \text{(WF-Store2)}$$

$$\frac{dom(\lambda) \subseteq dom(\Lambda_T) \quad \forall id \in dom(\lambda) \, . \, \cdot; \Sigma; \Lambda_T; \cdot \vdash \lambda(id) : \tau \wedge \tau \leq \Lambda_T(id)}{\Sigma; \Lambda_T \vdash \lambda} \quad \text{(WF-LocalLambda2)}$$

$$\frac{\Sigma; \Lambda_T \vdash \mu \quad \Sigma; \Lambda_T \vdash b \quad \Sigma; \Lambda_T \vdash \lambda \quad \exists \Gamma, \ell_c. \, \Gamma; \Sigma; \Lambda_T; \ell_c \vdash t : \tau}{\Sigma; \Lambda_T \vdash \langle t \mid \mu \mid b \mid \lambda \rangle^i} \quad \text{(WF-ProgramConfig)} \qquad \frac{\Sigma; \Lambda_T \vdash \langle t \mid \mu \mid b \mid \lambda \rangle^i \quad \Sigma; \Lambda_T \vdash P}{\Sigma; \Lambda_T \vdash \{\langle t \mid \mu \mid b \mid \lambda \rangle^i\} \cup P} \quad \text{(WF-Program2)}$$

$$\frac{\Sigma; \Lambda_T \vdash P \quad \Sigma; \Lambda_T \vdash M \quad \Sigma; \Lambda_T \vdash S \quad \Sigma; \Lambda_T \vdash \Lambda \quad dom(\Sigma) = range(\Lambda_T)}{\Sigma; \Lambda_T \vdash P \mid M \mid S \mid \Lambda} \quad \text{(WF-Config)}$$

**Figure 14** Selected well-formedness rules

## 5 Properties of CTRD

In this section, we discuss correctness and consistency properties of CTRD, respectively.

### 5.1 Correctness properties

We prove two correctness properties of CTRD: type soundness and noninterference. Type soundness is an essential property of a type system, and it guarantees that well-typed terms do not "go wrong". The proof consists of two parts: preservation and progress. The preservation theorem states that for a well-typed program and corresponding static and runtime environment, the reduction keeps the well-formedness of the environment. Moreover, the progress theorem states that for any well-typed program, it either is a value or can be evaluated.

The noninterference property of CTRD shows that the type system prevents the data flow from available computations to consistent computations. That is, any values that are labeled with ava will not influence the mutation of locations that are labeled with con. Due to this property, CTRD can be safely used to solve the problem in figure 1b.

The complete definitions, theorems and full proofs are in our companion technical report [30].

### 5.2 Consistency properties

In this section, we use abstract executions [1] to formalize consistency properties of CTRD.

#### 5.2.1 Abstract computation

**Definition 5.1** (Abstract executions). *Let $L = P \mid M \mid S \mid \Lambda$ be a well-formed configuration, i.e., $\Sigma; \Lambda_T \vdash P \mid M \mid S \mid \Lambda$ for some $\Sigma$ and $\Lambda_T$. An* abstract execution *for con operations is a tuple $\mathbb{A} = \langle L, OP, RB, RVAL, SP, VIS, AR \rangle$ where:*

**Consistency types for replicated data**

- $OP : \mathbb{V} \to \{op_\ell^\mathtt{v}(o, v)\}$ *maps events to operations.*
- $RVAL : \mathbb{V} \times$ Values $\cup \{\triangledown\}$ *describes the value returned by the operation, or the special symbol $\triangledown$ ($\triangledown \notin$ Values which means there is no return value for the operation).*
- $RB \subseteq \mathbb{V} \times \mathbb{V}$ *records the returns-before order.*
- $SP : Ids \to \mathbb{V}$ *maps client ids to events.*
- $VIS \subseteq \mathbb{V} \times \mathbb{V}$ *records whether the effects of an operation are visible to another operation on the server side. We assume that only events with read operations can observe the effect of other operations.*
- $AR \subseteq \mathbb{V} \times \mathbb{V}$ *records the arbitration order which is a total order of operations across all programs.*

The well-formedness of abstraction execution can be found in our technical report [30]. Rules in figure 15 provide an operational way to associate abstract executions with CTRD. Here we use the information from action $a$ as we mentioned in section 4.3. $a$ contains a label $\ell$ and an operation field op.

The remaining transitions of the system are considered as internal changes that do not affect history and are handled by rule (A-Internal).

The idea of constructing the abstraction computation is as follows.

A-READ applies to all the events with read operations. The read operation $rd_\ell^\mathtt{v}(o, v, R)$ contains the consistency level $\ell$ of the operation, the event id $\mathtt{v}$, and the value $v$ the operations returns from a server or a server set $R$. Since we assume that only they can observe the effect from the event, the VIS relation is updated to include the pair set where all the event happened locally or on $R$ should be visible to event $\mathtt{v}$.

A-WRITE-1 applies to all the events with con-labeled read or create reference operations. The operation $op_\mathsf{con}^\mathtt{v}(o, v)$ contains the consistency level con of the operation, the event id $\mathtt{v}$ and the value $v$ that is written to a location $o$. The AR relation is updated according to the event sequence recorded on each server-side.

A-WRITE-2 applies to all the events with ava-labeled read or create reference operations that happen locally. The operation $op_\mathsf{ava}^\mathtt{v}(o, v)$ contains the consistency level ava of the operation, the event id $\mathtt{v}$ and the value $v$ that is written to a location $o$. The effect of the operation is not directly applied to any remote servers. Thus relations related to the remote servers such as AR are not updated.

A-MSGPROCESS applies to all the events with ava-labeled read or create reference operations that occur remotely. This rule is a further propagation step for the events in A-WRITE-2 and will update the AR relation accordingly.

### 5.2.2 Guarantees of CTRD

The following example shows that the CTRD system exhibits a non-monotonic write anomaly if we consider all the operations in history.

**Example** (Non-monotonic write anomaly). *Consider the following system consisting of two programs.*

$$\frac{\begin{array}{c} L \overset{(\ell_c, rd_\ell^\mathbb{v}(o,v,R))}{\twoheadrightarrow} L' \qquad \mathbb{v} \notin dom(\text{OP}) \qquad \text{OP}' = \text{OP}[\mathbb{v} \mapsto rd_\ell^\mathbb{v}(o,v)] \\ \text{RB}' = \text{RB} \cup (\{\mathbb{w} \mid \mathbb{w} \in R.seq\} \cup \{\text{SP}(i)\} \times \{\mathbb{v}\}) \\ \text{SP}' = \text{SP}[i \mapsto \text{SP}(i) \cup \{\mathbb{v}\}] \qquad \text{VIS}' = \text{VIS} \cup (\{\mathbb{w} \mid \mathbb{w} \in R.seq\} \times \{\mathbb{v}\}) \cup (\{\text{SP}(i)\} \times \{\mathbb{v}\}) \\ \text{RVAL}' = \text{RVAL}[\mathbb{v} \mapsto v] \end{array}}{\langle L, \text{OP}, \text{RB}, \text{RVAL}, \text{SP}, \text{VIS}, \text{AR} \rangle \overset{(\ell_c, rd_\ell^\mathbb{v}(o,v,R))}{\twoheadrightarrow}_i \langle L', \text{OP}', \text{RB}', \text{RVAL}', \text{SP}', \text{VIS}', \text{AR} \rangle} \text{(A-Read)}$$

$$\frac{\begin{array}{c} P \mid M \mid S \mid \Lambda \overset{(\ell_c, \text{op}_{\text{con}}^\mathbb{v}(o,v))}{\twoheadrightarrow} P' \mid M \mid S' \mid \Lambda' \qquad \mathbb{v} \notin dom(\text{OP}) \qquad \text{OP}' = \text{OP}[\mathbb{v} \mapsto \text{op}_{\text{con}}^\mathbb{v}(o,v)] \\ \text{RB}' = \text{RB} \cup ((\{\mathbb{w} \mid \forall S_r \in S. \mathbb{w} \in S_r.seq\} \cup \{\text{SP}(i)\}) \times \{\mathbb{v}\}) \\ \text{SP}' = \text{SP}[i \mapsto \text{SP}(i) \cup \{\mathbb{v}\}] \qquad \text{AR}' = \text{AR} \cup (\{\mathbb{w} \mid \forall S_r \in S. \mathbb{w} \in S_r.seq\} \times \{\mathbb{v}\}) \\ \text{op} \in \{\text{wr}, \text{ref}\} \qquad \text{RVAL}' = \begin{cases} \text{RVAL}[\mathbb{v} \mapsto o] & \text{if op} = ref \\ \text{RVAL}[\mathbb{v} \mapsto \text{unit}] & \text{if op} = wr \end{cases} \end{array}}{\langle P \mid M \mid S \mid \Lambda, \text{OP}, \text{RB}, \text{RVAL}, \text{SP}, \text{VIS}, \text{AR} \rangle \overset{(\ell_c, \text{op}_{\text{con}}^\mathbb{v}(o,v))}{\twoheadrightarrow}_i \langle P' \mid M \mid S' \mid \Lambda', \text{OP}', \text{RB}', \text{RVAL}', \text{SP}', \text{VIS}, \text{AR}' \rangle} \text{(A-Write-1)}$$

$$\frac{\begin{array}{c} P \mid M \mid S \mid \Lambda \overset{(\ell_c, \text{op}_{\text{ava}}^\mathbb{v}(o,v))}{\twoheadrightarrow} P' \mid M \mid S \mid \Lambda \qquad \mathbb{v} \notin dom(\text{OP}) \qquad \text{OP}' = \text{OP}[\mathbb{v} \mapsto \text{op}_{\text{ava}}^\mathbb{v}(o,v)] \\ \text{RB}' = \text{RB} \cup \{\text{SP}(i)\}) \times \{\mathbb{v}\}) \qquad \text{SP}' = \text{SP}[i \mapsto \text{SP}(i) \cup \{\mathbb{v}\}] \\ \text{op} \in \{\text{wr}, \text{ref}\} \qquad \text{RVAL}' = \begin{cases} \text{RVAL}[\mathbb{v} \mapsto o] & \text{if op} = ref \\ \text{RVAL}[\mathbb{v} \mapsto \text{unit}] & \text{if op} = wr \end{cases} \end{array}}{\langle P \mid M \mid S \mid \Lambda, \text{OP}, \text{RB}, \text{RVAL}, \text{SP}, \text{VIS}, \text{AR} \rangle \overset{(\ell_c, \text{op}_{\text{ava}}^\mathbb{v}(o,v))}{\twoheadrightarrow}_i \langle P' \mid M \mid S \mid \Lambda, \text{OP}', \text{RB}', \text{RVAL}', \text{SP}', \text{VIS}, \text{AR} \rangle} \text{(A-Write-2)}$$

$$\frac{\begin{array}{c} P \mid M \mid S \mid \Lambda \overset{(\ell_c, \text{op}_{\text{ava}}^\mathbb{v}(o,v))}{\twoheadrightarrow} P \mid M' \mid S' \mid \Lambda' \qquad M = \{\text{update}(o, v, R, \mathbb{v})\} \bigcup M'' \qquad \text{op} \in \{\text{wr}, \text{ref}\} \\ \exists r \in R.\mathbb{v} \cdot r.seq \in \{s.seq \mid s \in S'\} \qquad \text{AR}' = \text{AR} \cup (\{\mathbb{w} \mid \mathbb{w} \in r.seq\} \times \{\mathbb{v}\}) \end{array}}{\langle P \mid M \mid S \mid \Lambda, \text{OP}, \text{RB}, \text{RVAL}, \text{SP}, \text{VIS}, \text{AR} \rangle \overset{(\ell_c, \text{op}_{\text{ava}}^\mathbb{v}(o,v))}{\twoheadrightarrow} \langle P \mid M' \mid S' \mid \Lambda', \text{OP}, \text{RVAL}, \text{RB}, \text{SP}, \text{VIS}, \text{AR}' \rangle} \text{(A-MsgProcess)}$$

$$\frac{P \mid M \mid S \mid \Lambda \overset{a}{\twoheadrightarrow} P' \mid M' \mid S \mid \Lambda \qquad a = (\ell_c, \varepsilon)}{\langle P \mid M \mid S \mid \Lambda, \text{OP}, \text{RB}, \text{RVAL}, \text{SP}, \text{VIS}, \text{AR} \rangle \overset{a}{\twoheadrightarrow}_i \langle P' \mid M' \mid S \mid \Lambda, \text{OP}, \text{RVAL}, \text{RB}, \text{SP}, \text{VIS}, \text{AR} \rangle} \text{(A-Internal)}$$

**Figure 15** Abstraction computation for CTRD.

**Consistency types for replicated data**

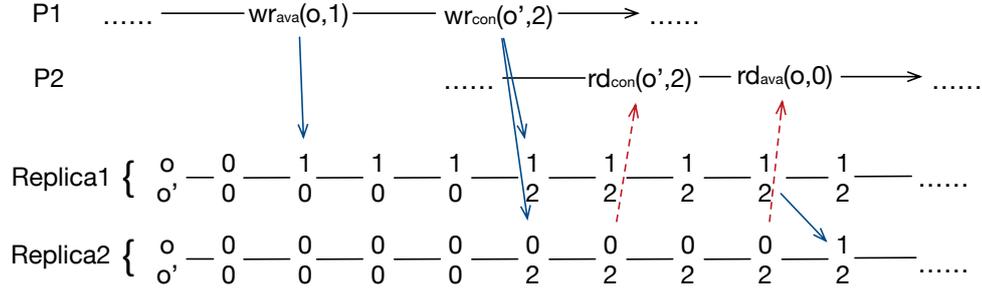

*Due to the slow propagation of the ava operation $wr_{ava}(o, 1)$, the visible order for a program's writes is not preserved anymore. The possible execution order violates sequential consistency.*

However, if we separate the analysis for ava and con operations, we can find more interesting properties. Due to the noninterference property, there is no information-flow from ava operations to con operations, which means ava operations do not affect con operations. Then we can analyze the consistency guarantees for con operations and ava operations separately.

Based on the collection of execution histories (see figure 15), the following theorem states that for well-formed programs, con operations provide sequential consistency.

**Notation** We clarify the following notations for the following theorem 5.1:
- $PAIRSET_{con} = \{(a, b) \mid (a, b) \in PAIRSET \wedge a = op_{con}^{v}(o, v) \wedge b = op_{con}^{v'}(o', v')\}$, $MAP_{con} = \{(a \mapsto MAP(a)) \mid a \in MAP.keys \wedge a = op_{con}^{v}(o, v)\}$.
- PO refers to program order: $PO = RB \cap SP$.
- Composition pairset $pair; pair' = \{(a, c) \mid \exists b \in A : a \xrightarrow{pair} b \xrightarrow{pair'} c\}$ where $a \xrightarrow{pair} b$ denotes $(a, b) \in pair$.
- Inverse pairset $pair^{-1} = \{(a, b) \mid b \xrightarrow{pair} a\}$.
- Negative pairset $\neg pair = \{(a, b) \mid a \not\xrightarrow{pair} b\}$.
- $\mathcal{F}$ is an abstract function for computing the return value of a operation, and its definition in CTRD is given as follows:
  $\mathcal{F}(rd_{\ell}^{v}(o, v, R)) = v, \mathcal{F}(wr_{\ell}^{v}(o, v)) = \text{unit}, \mathcal{F}(ref_{\ell}^{v}(o, v)) = o$

**Theorem 5.1** (Sequential consistency for con operations in CTRD)**.**
*Let $L = P \mid M \mid S \mid \Lambda$ be a well-formed configuration, i.e., $\Sigma; \Lambda_T \vdash P \mid M \mid S \mid \Lambda$ for some $\Sigma$ and $\Lambda_T$. If $\langle L, \emptyset, \emptyset, \emptyset, \emptyset, \emptyset, \emptyset \rangle \twoheadrightarrow_i^* \langle L', OP_{con}, RB_{con}, RVAL_{con}, SP_{con}, VIS_{con}, AR_{con} \rangle$ then*
- $PO_{con} \subseteq VIS_{con}$.
- $AR_{con}; VIS_{con} \subseteq VIS_{con}$ and $AR_{con}^{-1}; \neg VIS_{con} \subseteq \neg VIS_{con}$.
- $\forall e \in \mathbb{V} : RVAL_{con}(e) = \mathcal{F}(OP_{con}(e))$.

*Proof sketch.* By induction on the derivation of the abstraction execution with case analysis of the last applied rule. We can start with the assumed execution state where all three conditions hold. Then, we apply each rule in figure 15, and analyze the state of the next step. Note that we need to specify the definition of PO which is the program order: $PO_{con} = RB_{con} \cap SP_{con}$, and the update of PO can be simplified as $PO' = PO \cup (SP(i) \times \{\mathbb{v}\})$. See our companion technical report [30] for the complete proof. □

**Notation** We clarify the following notations for the following theorem 5.2:
- $PAIRSET_{ava} = \{(a,b) \mid (a,b) \in PAIRSET \land a = op_{ava}^{\mathbb{v}}(o,v) \land b = op_{ava}^{\mathbb{v}'}(o',v')\}$, $MAP_{ava} = \{(a \mapsto MAP(a)) \mid a \in MAP.keys \land a = op_{ava}^{\mathbb{v}}(o,v)\}$.
- We use $[x]$ to represent equivalence classes. The set of equivalence classes are denoted as $A/\approx_{pair}$.

**Theorem 5.2** (Eventual consistency for ava operations in CTRD).
*Let $L = P \mid M \mid S \mid \Lambda$ be a well-formed configuration, i.e., $\Sigma; \Lambda_T \vdash P \mid M \mid S \mid \Lambda$ for some $\Sigma$ and $\Lambda_T$. If $\langle L, \emptyset, \emptyset, \emptyset, \emptyset, \emptyset, \emptyset \rangle \twoheadrightarrow_i^* \langle L', OP_{ava}, RB_{ava}, RVAL_{ava}, SP_{ava}, VIS_{ava}, AR_{ava} \rangle$ then*

- $\forall e \in \mathbb{V} : \forall f \in E/\approx_{SP}: |\{e' \in [f] \mid (e \xrightarrow{RB} e') \land (e \xrightarrow{VIS} e')\}| < \infty$
- $\forall e \in \mathbb{V} : RVAL_{ava}(e) = \mathcal{F}(OP_{ava}(e))$.

*Proof sketch.* By analyzing rule (A-Msg-Process) we can see that all the available writes will be propagated to the server side, which will be visible for available reads after a finite number of steps; thus, eventual consistency is preserved. □

## 6 CTRD Extension: CTRD$^c$

CTRD prevents unexpected data flow, but it still has some limitations on efficiency. The section introduces CTRD$^c$, which extends CTRD with clone operations. We first explain the restrictions of CTRD and then introduce CTRD$^c$ as the solution.

**Motivation** Consider the following code.

```
1  let x = ref_con (3_con,(con, 1)) in
2  let y = ref_con (x, (con, 2)) in
3  let z = ref_con (y, (con, 3)) in
4  …
```

Using the semantics in CTRD, in Line 1 to 3, we generate consistent references $x$, $y$, and $z$, which requires three times synchronizations among the servers. The performance of the example is acceptable because the size is relatively small. The reference $z$ points to a reference graph that contains two nodes (reference $y$ and reference $x$). When we have a graph that contains 10, 100, 1000, or even more nodes, we need to reduce the synchronization costs to make the language practical.

We would like to introduce CTRD with the optimized upgrading operation clone to solve the problem with this motivation.

**Extended syntax in CTRD$^c$** The intention of CTRD$^c$ is to rewrite the example in section 6 as the following.

**Consistency types for replicated data**

$$
\begin{array}{rll}
\tau & ::= & \ldots \qquad\qquad\qquad\text{same types as CTRD} \\
& | & \{l_i : \tau_i{}^{i \in 1\ldots n}\}_\ell \quad \text{type of records} \\
t & ::= & \ldots \qquad\qquad\qquad\text{same expressions as CTRD} \\
& | & \{l_i = t_i{}^{i \in 1\ldots n}\}_\ell \quad \text{record} \\
& | & t.l \qquad\qquad\qquad\text{projection} \\
& | & \boxed{\text{clone}_\ell(t, id)} \quad\text{clone operation} \\
v & ::= & \ldots \qquad\qquad\qquad\text{same expressions as CTRD} \\
& | & \{l_i = v_i{}^{i \in 1\ldots n}\}_\ell \quad \text{record value}
\end{array}
$$

■ **Figure 16** CTRD$^c$: additional types, terms and values

```
1  let x = ref_loc (3_loc, (loc, 1)) in
2  let y = ref_loc (x, (loc, 2)) in
3  let z = ref_loc (y, (loc, 3)) in
4  let a = clone_con (z, (con, 4)) in
5  …
```

We see that the reference graph generation is made locally, and clone operation makes the remote version of the reference graph in one step. Thus, a significant amount of synchronization requirements are reduced, which is particularly essential for big graphs.

Figure 16 shows the syntactic extension of CTRD$^c$ from CTRD: First, we introduce a standard record type and record expressions, so that the program can be further developed as object-oriented style. Then we introduce a new expression clone$_\ell(t, id)$ for upgrading the type of expression $t$ from local to distributed label $\ell$ with identifier $id$.

**Semantics and Soundness of CTRD$^c$**   We include the extended semantics, preservation and progress theorems, and corresponding proofs in the companion technical report [30].

# 7 Related Work

**Consistency models**   According to the CAP theorem [7], (strong) consistency, availability, and partition tolerance cannot be achieved at the same time for any distributed system. Therefore, there exist multiple consistency models such as eventual consistency [3], causal consistency [15], read-after-write consistency [18], and sequential consistency [13]. They define different contracts between the system and the user. It is also common that a system provides a mixed level of consistency, such as fork consistency [17, 21], lazy replication [12], red-blue consistency [16], and observable atomic consistency [29, 31].

**Consistency types**   The concept of consistency types first appears in Disciplined Inconsistency [10]. The authors present the IPA storage system to provide consistency safety and error-bounded consistency. Our work instead uses language-based information-flow tracking techniques to guarantee noninterference, and we provide a formal semantics and type system. Further discussion of differences is included in section 3.3. ConSysT [5, 20] is another programming language that supports heterogeneous consistency specifications at the type level. However, the work is preliminary and the authors only present the syntax of a core calculus, as well as subtyping rules. Neither dynamic semantics nor correctness properties are given. Conflict-free replicated data types (CRDTs) [24] provide strong eventual consistency via commutative update operations. Cloud types [2] maintain availability via "read-my-own-writes" and achieve eventual consistency using GSP (global sequence protocol) [4] whenever the network is available. Mergeable replicated data types [11] are in the same spirit as CRDTs, but they use invertible relational specifications to automate the derivation of replicated versions of original, non-replicated data types.

**Information-flow control**   Information-flow control tracks how information propagates within a program to guarantee a specific property. It is a popular method in computer security, and research on information-flow security has been ongoing for over 40 years. Traditional security models such as the Bell-LaPadula model [8] developed mandatory access control where data is labeled using different security levels. Our work transfers the concept of security levels into consistency levels. Volpano, Irvine, and Smith [28] present one of the first papers with a type system for information-flow, and the work of Smith and Volpano [25] adds support for multiple threads. In contrast, our work develops a type system in a distributed setting. The style of our formalization is similar to [26] which provides a higher-order static security-typed language with references. The differences are discussed in section 3.3. We also obtain a noninterference property, which was first studied by Goguen and Meseguer [8]. LJGS [6] is another security type system for object-oriented languages. EnerJ [23] considers the isolation of two program parts under an energy-accuracy trade-off scenario. Recently, some works have studied the impact of weak memory models on information-flow security. Vaughan and Millstein [27] first studied noninterference under TSO (total store order). Mantel, Perner, and Sauer [19] extended this study to four different memory models.

# 8   Conclusion

In order to meet increasing demands for scalability, availability, and fault tolerance, distributed system developers need to trade consistency for availability. However, mixing strongly consistent and weakly consistent data within the same application can give rise to bugs that are difficult to find and fix. In this paper, we presented a type system that enables the safe use of both kinds of data within the same program. The proof of a noninterference theorem guarantees that updates of weakly-consistent data can never affect strongly-consistent data. The type system also guarantees sequential consistency for so-called "con" operations. We also extended the core type system with label upgrading and with flexible consistency choices in order to support more application scenarios.

**Consistency types for replicated data**

We believe this result is an important step towards programming languages that enable the safe use of weakly consistent data alongside strongly consistent data in order to increase the scalability, availability, and fault tolerance of future distributed systems.

**Acknowledgements** This work is supported by the China Scholarship Council 201600160040, and by the Tianhe Supercomputer Project 2018YFB0204301.

**Consistency types for replicated data**[20]   Alessandro Margara and Guido Salvaneschi. "Consistency Types for Safe and Efficient Distributed Programming". In: *Proceedings of the 19th Workshop on Formal Techniques for Java-like Programs*. FTfJP. 2017, 8:1–8:2. ISBN: 978-1-4503-5098-3. DOI: 10.1145/3103111.3104044.

[21]   David Maziéres and Dennis Shasha. "Building Secure File Systems out of Byzantine Storage". In: *Proceedings of the Twenty-First Annual Symposium on Principles of Distributed Computing*. PODC. 2002, pages 108–117. ISBN: 1-58113-485-1. DOI: 10.1145/571825.571840.

[22]   Diego Ongaro and John K. Ousterhout. "In Search of an Understandable Consensus Algorithm". In: *2014 USENIX Annual Technical Conference (USENIX ATC 14)*. USENIX ATC. 2014, pages 305–319. ISBN: 978-1-931971-10-2. URL: https://www.usenix.org/system/files/conference/atc14/atc14-paper-ongaro.pdf.

[23]   Adrian Sampson, Werner Dietl, Emily Fortuna, Danushen Gnanapragasam, Luis Ceze, and Dan Grossman. "EnerJ: approximate data types for safe and general low-power computation". In: *Proceedings of the 32nd ACM SIGPLAN Conference on Programming Language Design and Implementation, PLDI 2011, San Jose, CA, USA, June 4-8, 2011*. PLDI. 2011, pages 164–174. ISBN: 978-1-4503-0663-8. DOI: 10.1145/1993316.1993518.

[24]   Marc Shapiro, Nuno M. Preguiça, Carlos Baquero, and Marek Zawirski. "Conflict-Free Replicated Data Types". In: *Stabilization, Safety, and Security of Distributed Systems SSS*. 2011, pages 386–400. DOI: 10.1007/978-3-642-24550-3_29.

[25]   Geoffrey Smith and Dennis Volpano. "Secure Information Flow in a Multi-Threaded Imperative Language". In: *Proceedings of the 25th ACM SIGPLAN-SIGACT Symposium on Principles of Programming Languages*. POPL. New York, NY, USA, 1998, pages 355–364. ISBN: 0-89791-979-3. DOI: 10.1145/268946.268975.

[26]   Matías Toro, Ronald Garcia, and Éric Tanter. "Type-Driven Gradual Security with References". In: *ACM Transactions on Programming Languages and Systems* 40.4 (2018), 16:1–16:55. ISSN: 0164-0925. DOI: 10.1145/3229061.

[27]   Jeffrey A. Vaughan and Todd D. Millstein. "Secure Information Flow for Concurrent Programs under Total Store Order". In: *25th IEEE Computer Security Foundations Symposium, CSF 2012, Cambridge, MA, USA, June 25-27, 2012*. CSF. 2012, pages 19–29. DOI: 10.1109/CSF.2012.20.

[28]   Dennis Volpano, Cynthia Irvine, and Geoffrey Smith. "A Sound Type System for Secure Flow Analysis". In: *Journal of Computer Security* 4.2–3 (Jan. 1996), pages 167–187. ISSN: 0926-227X. DOI: 10.3233/JCS-1996-42-304.

[29]   Xin Zhao and Philipp Haller. "Observable atomic consistency for CvRDTs". In: *Proceedings of the 8th ACM SIGPLAN International Workshop on Programming Based on Actors, Agents, and Decentralized Control*. AGERE!@SPLASH. 2018, pages 23–32. ISBN: 978-1-4503-6066-1. DOI: 10.1145/3281366.3281372.

[30]   Xin Zhao and Philipp Haller. *On consistency types for lattice-based distributed programming languages*. Technical report. 2019. arXiv: 1907.00822. URL: http://arxiv.org/abs/1907.00822.6:28

**Consistency types for replicated data**

**About the authors**

**Xin Zhao** is a PhD student in the School of Electrical Engineering and Computer Science at KTH Royal Institute of Technology in Stockholm, Sweden. She currently works on programming languages, distributed programming and type systems. For more information, see https://people.kth.se/~xizhao/. Contact her at xizhao@kth.se.

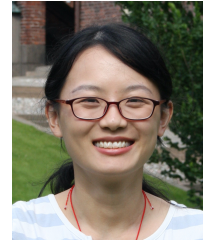

**Philipp Haller** is an Associate Professor of Computer Science at KTH Royal Institute of Technology in Stockholm, Sweden. He was part of the team that received the 2019 ACM SIGPLAN Programming Languages Software Award for the development of the Scala programming language. He received a Ph.D. from École Polytechnique Fédérale de Lausanne, EPFL, Switzerland and a Diplom-Informatiker degree from Karlsruhe Institute of Technology, Germany. His main research interests are programming language design and implementation, type systems, concurrency, and distributed programming. For more information, see https://people.kth.se/~phaller/. Contact him at hallerp@acm.org.

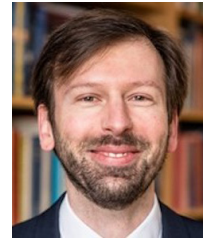